# Quasi-Fractal UCA Based $N$-Dimensional OAM Orthogonal Transmission

Hongyun Jin†, Wenchi Cheng†, and Wei Zhang‡

†State Key Laboratory of Integrated Services Networks, Xidian University, Xi'an, China
‡School of Electrical Engineering and Telecommunications, The University of New South Wales, Sydney, Australia
E-mail: {*hongyunjin@stu.xidian.edu.cn*, *wccheng@xidian.edu.cn*, *w.zhang@unsw.edu.au*}

*Abstract*—The vortex electromagnetic wave carried by multiple orthogonal orbital angular momentum (OAM) modes in the same frequency band can be applied to the field of wireless communications, which greatly increases the spectrum efficiency. The uniform circular array (UCA) structure is widely used to generate or receive vortex electromagnetic waves with multiple OAM-modes. However, the maximum number of orthogonal OAM-modes based on UCA is usually limited to the number of array-elements of the UCA antenna, leaving how to utilize more OAM-modes to achieve higher spectrum efficiency given a fixed number of array-elements as an intriguing question. In this paper, we propose an $N$-dimensional quasi-fractal UCA ($N$D QF-UCA) antenna structure in different fractal geometry layouts to break through the limits of array-elements number on OAM-modes number. We develop the $N$-dimensional OAM modulation ($N$OM) and demodulation ($N$OD) schemes for OAM multiplexing transmission with the OAM-modes number exceeding the array-elements number, which is beyond the traditional concept of multiple antenna based wireless communications. Then, we investigate different dimensional multiplex transmission schemes based on the corresponding QF-UCA antenna structure with various array-elements layouts. Simulation results show that our proposed schemes can obtain a higher spectrum efficiency.

## I. Introduction

FOR the next-generation 6G wireless system application scenarios, there are requirements for enhanced channel capacity and higher spectrum efficiency (SE) [1]. Vortex electromagnetic wave technology, as a new mode division multiplexing technology, has received significant researech attention in the field of wireless communications [2]. Utilizing mutually orthogonal multimodal vortex electromagnetic wave transmission techniques, the performance of communication systems can be greatly improved without increasing the current spectrum bandwidth, thus leading to a significant increase in the SE [3].

The uniform circular array (UCA) is the classical method for generating or receiving vortex waves. UCA antenna-based OAM offers notable advantages in terms of flexibility and convenience in generating OAM-modes, thus prompting a wide investigation and high expectations of its integration within wireless communications. The authors of [4] generated OAM beams with UCA fed by in-phase and verified the generation of different OAM-modes on an experimental platform. The authors of [5] proposed a radial UCA for OAM generation and dual-mode communication based on a multilayer design and gave a theoretical derivation of radial UCA for OAM generation. The authors of [6] investigated the OAM-mode of vortical radio wave generated by UCA, derived theoretical equations for the radiation field, and determined the factors affecting the pattern distribution. Then, an experimental setup of a vortex radio beam was created and the radiation field was measured to verify the theoretical results.

The design of the vortex electromagnetic wave multiplexing transmission scheme based on multi-array antenna holds significant importance in enhancing the channel capacity [7]. The authors of [8] proposed the OAM-embedded-MIMO communication system with UCA antenna to obtain the SE gain for joint OAM and massive-MIMO based wireless communications. The authors of [9] proposed a general scheme for multi-carrier and multi-mode OAM communication based on UCAs, which reduces the burden of estimating channel matrices to better realize the transmission and reception of vortex electromagnetic waves. The authors of [10] proposed a small-scale circular phased array antenna for OAM-carrying radio beams. Meanwhile, the generation of multiple pure or mixed OAM beams with helical phase fronts is also presented, where the superposition of multiple OAM states provides more possibilities to enhance the channel capacity.

However, in order to increase the number of available OAM-modes, conventional vortex electromagnetic waves typically adopt higher-order OAM-modes, which leads to a few problems. On the one hand, higher-order OAM-modes require more transmitter antenna array-elements and larger transmitter antenna aperture. On the other hand, due to the hollow nature of the vortex electromagnetic wave, the vortex electromagnetic wave of the higher-order OAM-modes disperses seriously [11]. How to effectively utilize the aperture of the transmitter antenna to design the layouts of array-elements for transmitting vortex electromagnetic wave represents a significant research challenge. This endeavor aims to highlight the great advantage of the vortex electromagnetic wave over the traditional plane wave, receiving considerable research attention.

In order to solve the above problems, we design a new antenna geometry layout and provide an $N$-dimensional OAM multiplexing method to generate more orthogonal OAM-modes, where the number of OAM-modes is greater than that of array-elements. First, we design an $N$-dimensional ($N$D) quasi-fractal UCA (QF-UCA) antenna layout in which each dimensional QF-UCA is taken into account as an integral unit. There are shared array-elements between two adjacent UCAs in QF-UCA, which can generate or receive different OAM-modes in the two adjacent UCAs, respectively. The utilization of shared array-elements is able to break through the limitations of the number of array-elements on the number of OAM-modes, and provide a new solution to increase the number of available OAM-modes when the antenna aperture and the number of array-elements are fixed. Then, we develop $N$-dimensional OAM modulation ($N$OM) and demodulation ($N$OD) schemes based on discrete Fourier transformation (DFT) for multiple OAM-modes transmission. Vortex electromagnetic waves for high-capacity transmission can adopt the multi-dimensional modulation and demodulation scheme to further multiplex lower-order modes, which is a novel multiplexing transmission scheme [12]. Simulation results show that our proposed schemes can obtain a higher SE.

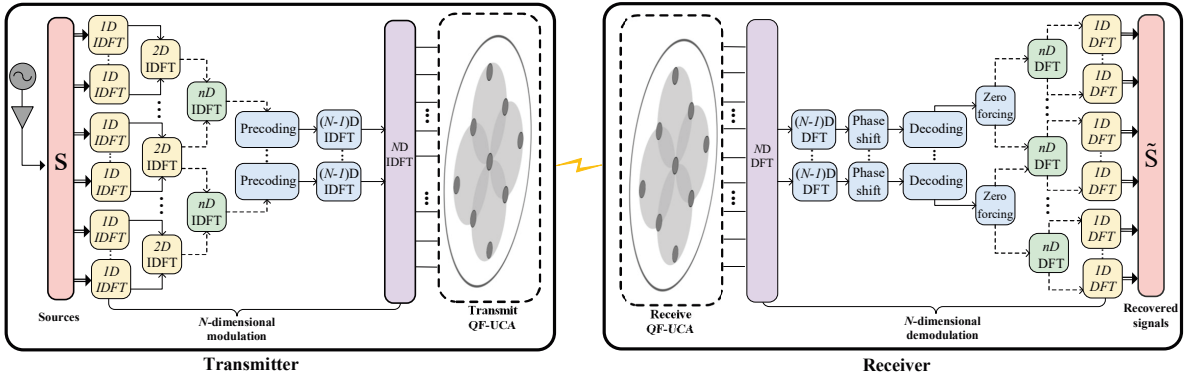

Fig. 1. The system model for $N$-dimensional OAM multiplexing transmission with QF-UCA.

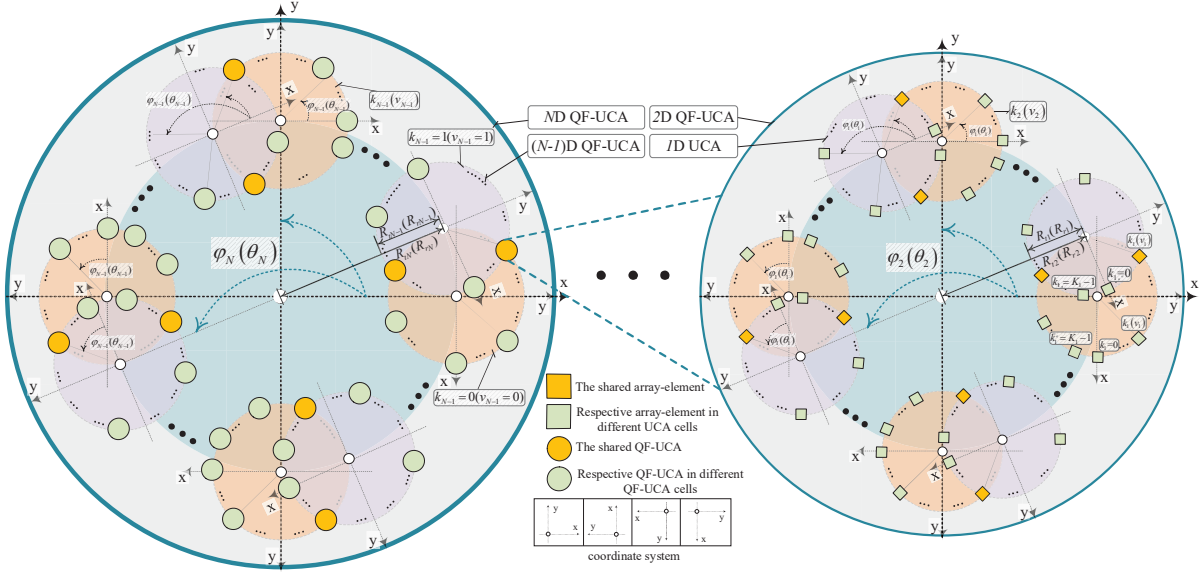

Fig. 2. The $N$D QF-UCA structure.

The rest of this paper is organized as follows. In Section II, the system model and the structure of $N$D QF-UCA antenna are provided. In Section III, the $N$OM and $N$OD schemes and the channel model are analyzed. Section IV provides the numerical results. Finally, we conclude this paper in Section V.

Notation: Matrices and vectors are denoted by the capital letters and lowercase letters in bold, respectively. The notation blkdiag ($\boldsymbol{X}$) represents a diagonal block matrix with $\boldsymbol{X}$ as its main diagonal elements. The notations $(\cdot)^H$ and $(\cdot)^T$ denote the Hermitian and the transpose of a matrix or vector, respectively.

## II. THE SYSTEM MODEL FOR $N$D OAM MULTIPLEXING TRANSMISSION BASED ON $N$D QF-UCA

Fig. 1 depicts the system model for $N$-dimensional OAM multiplexing transmission with QF-UCA antennas. The transmitter includes $n$D IDFT ($n \in [1, N]$) modulation, precoding, and transmit $N$D QF-UCA antenna. The 1D, $\cdots$, and $n$D IDFT modulation forms the $N$-dimensional IDFT based OAM modulation for input signals. The receiver includes receive $N$D QF-UCA antenna, $n$D DFT ($n \in [1, N]$) demodulation, phase shift, decoding, and zero forcing. The 1D, $\cdots$, and $n$D DFT demodulation forms the $N$-dimensional DFT based OAM demodulation for received signals. The precoding and decoding are designed together to ensure that the OAM-modes are orthogonal at the receiver. The phase shift converts the antisymmetric elements of the equivalent channel matrix into symmetric elements, then decoding and zero forcing are performed. The signals carrying different OAM-modes are recovered after $N$D OAM demodulation based on the orthogonality between the OAM-modes.

Fig. 2 depicts the $N$D QF-UCA structure. The transmit $N$D QF-UCA antenna has $N_t$ array-elements, which $n$D QF-UCA ($n \in [1, N]$) cells are divided into $K_n$ ($n$–1)D QF-UCA cells. Since there exist shared array-elements among different UCA cells, the total number of transmitted data streams is larger than the number of array-elements equipped in the $N$D QF-UCA, resulting in $\prod_{n=1}^{N} K_n > N_t$. The receive $N$D QF-UCA antenna has $N_r$ array-elements, which $n$D QF-UCA ($n \in [1, N]$) cells are divided into $V_n$ ($n$–1)D QF-UCA cells. The $N$D QF-UCA antenna can receive $\prod_{n=1}^{N} V_n$ ($\prod_{n=1}^{N} V_n > N_r$) data streams. The radius of the $n$D ($n \in [1, N]$) QF-UCA is the distance from the center of the $n$D QF-UCA to the center of the ($n$–1)D QF-UCA. We denote by $R_{tn}$ and $R_{rn}$ the radii of $n$D transmit and receive QF-UCA cells, respectively. Then, we denote by $R_{tE}$ and $R_{rE}$ the radius of the entire transmit and receive $N$D QF-UCA, respectively, where $R_{tE} = \sum_{n=1}^{N} R_{tn}$ and $R_{rE} = \sum_{n=1}^{N} R_{rn}$. Generally, we assume that $R_E = R_{tE} = R_{rE}$.

The $n$D QF-UCA cell or array-element indexed with $k_n$ ($k_n \in [0, K_n - 1]$, $n \in [1, N]$) at the transmitter and those indexed with $v_n$ ($v_n \in [0, V_n - 1]$, $n \in [1, N]$) at the receiver are uniformly distributed along the center of $n$D QF-UCA, respectively. In the $n$D QF-UCA coordinate system, $x$-axis is set as the direction from the center of $n$D QF-UCA cell to the first ($n$–1)D QF-UCA cell ($k_n = 0$, $v_n = 0$) while $z$-axis is the centering normal line pointing to the directly opposite receive QF-UCA cell. Correspondingly, $y$-axis are decided by the right-hand spiral rule. Then, we denote by ($k_N$, $\cdots$, $k_1$) and ($v_N$, $\cdots$, $v_1$) the global index of the array-element of the entire transmit and receive $N$D QF-UCA antenna, respectively. The parameters $\varphi_n = \frac{2\pi k_n}{K_n}$ and $\theta_n = \frac{2\pi v_n}{V_n}$ denote the azimuths within the $n$D QF-UCA cell at the transmitter and receiver, respectively. In this paper, we assume that the transmit and receive $n$D QF-UCA antennas are strictly aligned with each other and that the design-related requirements for antennas, such as the impedance matching and the spatial correlation, are well satisfied. Also, we mainly consider LOS channel in this paper.

## III. N-DIMENSIONAL OAM MULTIPLEXING TRANSMISSION SCHEME

### A. N-Dimensional IDFT Based OAM Modulation (NOM)

We denote by one-dimensional (1D) UCA the traditional single-loop UCA. The 1D IDFT based OAM modulation signal vector, represented by $\boldsymbol{X}_1 = \boldsymbol{W}_1 \boldsymbol{s}_1$, where $\boldsymbol{s}_1 = [s_{1\_0}, \cdots s_{1\_l_1}, \cdots s_{1\_(K_1-1)}]^T$, $s_{1\_l_1}$ is the signal corresponding to the OAM-mode $l_1$ ($l_1 \in [0, K_1 - 1]$) carried by 1D UCA, and $K_1$ is the number of array-elements on the 1D UCA. The notation $\boldsymbol{W}_1$ denotes the 1D IDFT modulation matrix of order $K_1$, can be given as follows:

$$\boldsymbol{W}_1 = \begin{bmatrix} 1 & \cdots & 1 & \cdots & 1 \\ \vdots & \ddots & \vdots & \ddots & \vdots \\ 1 & \cdots & e^{j\frac{2\pi l_1 k_1}{K_1}} & \cdots & e^{j\frac{2\pi (K_1-1)k_1}{K_1}} \\ \vdots & \ddots & \vdots & \ddots & \vdots \\ 1 & \cdots & e^{j\frac{2\pi l_1(K_1-1)}{K_1}} & \cdots & e^{j\frac{2\pi (K_1-1)^2}{K_1}} \end{bmatrix}. \quad (1)$$

Then, we expand to higher-dimensional OAM modulation based on higher-dimensional QF-UCA. We define by $\boldsymbol{E}_{n-1}$ the identity matrix of order $\prod_{i=1}^{n-1} K_i$, where the order is related to the number of each dimensional QF-UCA cells. Then, the $n$D IDFT modulation matrix of order $\prod_{i=1}^{n} K_i$ denoted by $\boldsymbol{W}_n$, can be given as follows:

$$\boldsymbol{W}_n = \begin{bmatrix} \boldsymbol{E}_{n-1} & \cdots & \boldsymbol{E}_{n-1} & \cdots & \boldsymbol{E}_{n-1} \\ \vdots & \ddots & \vdots & \ddots & \vdots \\ \boldsymbol{E}_{n-1} & \cdots & e^{j\frac{2\pi l_n k_n}{K_n}}\boldsymbol{E}_{n-1} & \cdots & e^{j\frac{2\pi (K_n-1)k_n}{K_n}}\boldsymbol{E}_{n-1} \\ \vdots & \ddots & \vdots & \ddots & \vdots \\ \boldsymbol{E}_{n-1} & \cdots & e^{j\frac{2\pi l_n(K_n-1)}{K_n}}\boldsymbol{E}_{n-1} & \cdots & e^{j\frac{2\pi (K_n-1)^2}{K_n}}\boldsymbol{E}_{n-1} \end{bmatrix}. \quad (2)$$

We denote by $\boldsymbol{W}_{k_n}$ the $n$D nested IDFT modulation matrix of order $\prod_{i=1}^{n} K_i$ with elements are related to the ($n$–1)D nested IDFT modulation matrix $\boldsymbol{W}_{k_{n-1}}$, can be given as follows:

$$\begin{aligned} \boldsymbol{W}_{k_n} &= \boldsymbol{W}_n \boldsymbol{\Lambda}_n \\ &= \begin{bmatrix} \boldsymbol{W}_{k_{n-1}} & \cdots & \boldsymbol{W}_{k_{n-1}} & \cdots & \boldsymbol{W}_{k_{n-1}} \\ \vdots & \ddots & \vdots & \ddots & \vdots \\ \boldsymbol{W}_{k_{n-1}} & \cdots & e^{j\frac{2\pi l_n k_n}{K_n}}\boldsymbol{W}_{k_{n-1}} & \cdots & e^{j\frac{2\pi (K_n-1)k_n}{K_n}}\boldsymbol{W}_{k_{n-1}} \\ \vdots & \ddots & \vdots & \ddots & \vdots \\ \boldsymbol{W}_{k_{n-1}} & \cdots & e^{j\frac{2\pi l_n(K_n-1)}{K_n}}\boldsymbol{W}_{k_{n-1}} & \cdots & e^{j\frac{2\pi (K_n-1)^2}{K_n}}\boldsymbol{W}_{k_{n-1}} \end{bmatrix}, \end{aligned} \quad (3)$$

where $\boldsymbol{\Lambda}_n = blkdiag(\boldsymbol{W}_{k_{n-1}}, \cdots, \boldsymbol{W}_{k_{n-1}})$ is the scalar block matrix of order $\prod_{i=1}^{n} K_i$.

Thus, with $\boldsymbol{s}_{n\_l_n} = [s_{n-1\_0} \cdots, s_{n-1\_l_{n-1}} \cdots, s_{n-1\_K_{n-1}-1}]^T$ ($n \in (1, N]$), the transmit signal vector on the cell indexed with $k_n$ on the $n$D QF-UCA, represented by $\boldsymbol{x}_{n\_k_n}$, can be derived as follows:

$$\boldsymbol{x}_{n\_k_n} = \sum_{l_n=0}^{K_n-1} \frac{e^{j\frac{2\pi l_n k_n}{K_n}}}{\sqrt{K_n}} \boldsymbol{W}_{k_{n-1}} \boldsymbol{s}_{n\_l_n}. \quad (4)$$

Then, distinguishing the nested IDFT modulation matrix from other dimensions, the ($N$–1)D nested IDFT modulation matrix is given as $\boldsymbol{W}_{k_{N-1}} = \boldsymbol{W}_{N-1} \boldsymbol{Q}_{N-1} \boldsymbol{\Lambda}_{N-1}$, where $\boldsymbol{W}_{N-1}$ is the ($N$–1)D IDFT modulation matrix of order $\prod_{n=1}^{N-1} K_n$ and $\boldsymbol{Q}_{N-1}$ is precoding matrix.

Then, we give the relevant signals in the $N$D QF-UCA. The modulated signal on array-element indexed with ($k_N$, $\cdots$, $k_1$) on the $N$D QF-UCA, represented by $x_{(k_N,\cdots,k_1)}$, can be given as follows:

$$x_{(k_N,\cdots,k_1)} = \sum_{l_N=0}^{K_N-1} \cdots \sum_{l_1=0}^{K_1-1} \frac{s_{(l_N,\cdots,l_1)}}{\sqrt{K_N \cdots K_1}} e^{j\frac{2\pi l_1 k_1}{K_1}} \cdots e^{j\frac{2\pi l_N k_N}{K_N}}, \quad (5)$$

where $s_{(l_N,\cdots,l_1)}$ is the transmit signal corresponding to $N$D QF-UCA OAM-mode $l_N$ ($l_N \in [0, K_N - 1]$), $\cdots$, $n$D QF-UCA OAM-mode $l_n$ ($l_n \in [0, K_n - 1]$), $\cdots$, and 1D UCA OAM-mode $l_1$.

Then, with $\boldsymbol{S}_N = [\boldsymbol{s}_{N\_0}, \cdots, \boldsymbol{s}_{N\_l_N}, \cdots, \boldsymbol{s}_{N\_K_N-1}]^T$, the transmit signal vector for $N$D QF-UCA, represented by $\boldsymbol{X}_N$, can be derived as follows:

$$\boldsymbol{X}_N = \boldsymbol{W}_N \begin{bmatrix} \boldsymbol{W}_{k_{N-1}} \boldsymbol{s}_{N\_0} \\ \vdots \\ \boldsymbol{W}_{k_{N-1}} \boldsymbol{s}_{N\_l_N} \\ \vdots \\ \boldsymbol{W}_{k_{N-1}} \boldsymbol{s}_{N\_K_N-1} \end{bmatrix} = \boldsymbol{W}_N \boldsymbol{\Lambda}_N \boldsymbol{S}_N. \quad (6)$$

### B. Wireless Channel Of $N$-Dimensional OAM Transmission

It is generally assumed that each dimensional QF-UCA cell corresponding to the transmitter and receiver satisfies $K_n = V_n$ ($n \in [1, N]$). In this paper, transmit signals are propagated in a LoS path. According to the path loss of radio waves in free space, the expression of the complex channel gain between the array-element indexed with ($k_N, \cdots, k_1$) on transmitter and the array-element indexed with ($v_N, \cdots, v_1$) on receiver, denoted by $h_{(v_N,\cdots,v_1)}^{(k_N,\cdots,k_1)}$, can be given as follows:

$$h_{(v_N,\cdots,v_1)}^{(k_N,\cdots,k_1)} = \frac{\beta\lambda}{4\pi} \cdot \frac{e^{-j\frac{2\pi}{\lambda} d_{(v_N,\cdots,v_1)}^{(k_N,\cdots,k_1)}}}{d_{(v_N,\cdots,v_1)}^{(k_N,\cdots,k_1)}}, \quad (7)$$

where $\lambda$ represents the radio wavelength, $\beta$ denotes the parameter gathering relevant constants on antenna array-elements. The notation $d_{(v_N,\cdots,v_1)}^{(k_N,\cdots,k_1)}$ is the transmission distance from the array-element indexed with ($k_N, \cdots, k_1$) on the transmitter to that indexed with ($v_N, \cdots, v_1$) on the receiver, as derived in Eq. (8), where $D$ denotes the vertical distance between transmitter and receiver.

We denote by $\boldsymbol{H}_{(v_N,\cdots,v_n)}^{(k_N,\cdots,k_n)}$ the $(n-1)$D wireless channel gain matrix between the transmitter $(n-1)$D QF-UCA indexed with ($k_N, \cdots, k_n$) and receiver $(n-1)$D QF-UCA indexed with ($v_N, \cdots, v_n$), respectively. Thus, the $n$D wireless channel gain matrix between the transmitter $n$D QF-UCA and receiver $n$D QF-UCA, denote by $\boldsymbol{H}_{(v_N,\cdots,v_{n+1})}^{(k_N,\cdots,k_{n+1})}$, can be given as follows:

$$\boldsymbol{H}_{(v_N,\cdots,v_{n+1})}^{(k_N,\cdots,k_{n+1})} = \begin{bmatrix} \boldsymbol{H}_{(v_N,\cdots,0)}^{(k_N,\cdots,0)} & \cdots & \boldsymbol{H}_{(v_N,\cdots,0)}^{(k_N,\cdots,K_n-1)} \\ \vdots & \ddots & \vdots \\ \boldsymbol{H}_{(v_N,\cdots,v_n)}^{(k_N,\cdots,0)} & \cdots & \boldsymbol{H}_{(v_N,\cdots,v_n)}^{(k_N,\cdots,K_n-1)} \\ \vdots & \ddots & \vdots \\ \boldsymbol{H}_{(v_N,\cdots,V_n-1)}^{(k_N,\cdots,0)} & \cdots & \boldsymbol{H}_{(v_N,\cdots,V_n-1)}^{(k_N,\cdots,K_n-1)} \end{bmatrix}. \quad (9)$$

Then, we denote by $\boldsymbol{H}_{(v_N)}^{(k_N)}$ the $(N{-}1)$D wireless channel gain matrix between the transmitter $(N{-}1)$D QF-UCA and receiver $(N{-}1)$D QF-UCA indexed with $k_N$ and $v_N$, respectively. Thus, the $N$D wireless channel gain matrix between transmitter and receiver, denote by $\boldsymbol{H}_N$, can be given as follows:

$$\boldsymbol{H}_N = \begin{bmatrix} \boldsymbol{H}_{(0)}^{(0)} & \cdots & \boldsymbol{H}_{(0)}^{(k_N)} & \cdots & \boldsymbol{H}_{(0)}^{(K_N-1)} \\ \vdots & \ddots & \vdots & \ddots & \vdots \\ \boldsymbol{H}_{(v_N)}^{(0)} & \cdots & \boldsymbol{H}_{(v_N)}^{(k_N)} & \cdots & \boldsymbol{H}_{(v_N)}^{(K_N-1)} \\ \vdots & \ddots & \vdots & \ddots & \vdots \\ \boldsymbol{H}_{(V_N-1)}^{(0)} & \cdots & \boldsymbol{H}_{(V_N-1)}^{(k_N)} & \cdots & \boldsymbol{H}_{(V_N-1)}^{(K_N-1)} \end{bmatrix}, \quad (10)$$

which is circulant block matrix, as indicated by the uniform circular cell structure of the $N$D QF-UCA at both the transmitter and receiver.

### C. $N$-Dimensional DFT Based OAM Demodulation(NOD)

We can derive the received signal vector $\boldsymbol{R}_N$ corresponding to $N$D QF-UCA cells through the LOS channels as follows:

$$\boldsymbol{R}_N = \boldsymbol{H}_N \boldsymbol{W}_N \boldsymbol{\Lambda}_N \boldsymbol{S}_N. \quad (11)$$

The signal vector after the $N$D OAM demodulation based on $N$D QF-UCA, denoted by $\boldsymbol{\mathscr{R}}_N$, can be derived as follows:

$$\boldsymbol{\mathscr{R}}_N = \boldsymbol{W}_N^H \boldsymbol{R}_N = \boldsymbol{W}_N^H \boldsymbol{H}_N \boldsymbol{W}_N \boldsymbol{\Lambda}_N \boldsymbol{S}_N = \boldsymbol{\mathscr{H}}_N \boldsymbol{\Lambda}_N \boldsymbol{S}_N, \quad (12)$$

where $\boldsymbol{W}_N^H$ is defined as the $N$D DFT demodulation matrix. For the circulant block matrix $\boldsymbol{H}_N$, there is a good property in connection with unitary decomposition that the unitary similarity transformation for $\boldsymbol{H}_N$ produces a diagonal block matrix. Specifically, using IDFT matrix $\boldsymbol{W}_N$ and DFT matrix $\boldsymbol{W}_N^H$, we have $\boldsymbol{W}_N^H \boldsymbol{H}_N \boldsymbol{W}_N = \boldsymbol{\mathscr{H}}_N$, where $\boldsymbol{\mathscr{H}}_N$ is a diagonal block matrix with $n$th element on the diagonal is $\boldsymbol{\mathscr{H}}_{N-1}$. We name the above-mentioned operation the $N$D OAM demodulation based on $N$D QF-UCA.

Then, performing a dimensionality reduction operation on Eq. (12). We take the $k_N$th row of $\boldsymbol{\mathscr{R}}_N$ and denote as $\tilde{\boldsymbol{R}}_{N-1}$, which denotes the signal vector corresponding to the $(N{-}1)$D QF-UCA cells and can be given as follows:

$$\tilde{\boldsymbol{R}}_{N-1} = \boldsymbol{\mathscr{H}}_{N-1} \boldsymbol{W}_{k_{N-1}} \tilde{\boldsymbol{S}}_{N-1} = \boldsymbol{\mathscr{H}}_{N-1} \boldsymbol{W}_{N-1} \boldsymbol{Q}_{N-1} \boldsymbol{\Lambda}_{N-1} \tilde{\boldsymbol{S}}_{N-1}. \quad (13)$$

The signal vector after the $(N{-}1)$D OAM demodulation based on $(N{-}1)$D QF-UCA, denoted by $\boldsymbol{\mathscr{R}}_{N-1}$, can be derived as follows:

$$\boldsymbol{\mathscr{R}}_{N-1} = \boldsymbol{U}_{N-1}^H \boldsymbol{P}_{N-1} \boldsymbol{W}_{N-1}^H \tilde{\boldsymbol{R}}_{N-1}, \quad (14)$$

where $\boldsymbol{W}_{N-1}^H$ is defined as the $(N{-}1)$D DFT demodulation matrix. $\boldsymbol{P}_{N-1}$ is the phase shift matrix, which converts the antisymmetric elements of the equivalent channel matrix into

$$d_{(v_N,\cdots,v_1)}^{(k_N,\cdots,k_1)} = \left\{ D^2 + [(R_{rN}\cos\theta_N + \cdots + R_{r1}\cos\theta_1) - (R_{tN}\cos\varphi_N + \cdots + R_{t1}\cos\varphi_1)]^2 \right.$$
$$\left. + [(R_{rN}\sin\theta_N + \cdots + R_{r1}\sin\theta_1) - (R_{tN}\sin\varphi_N + \cdots + R_{t1}\sin\varphi_1)]^2 \right\}^{\frac{1}{2}}$$
$$\approx D + \frac{1}{D} \sum_{n=1}^{N} [R_n^2(1 - \cos(\theta_n - \varphi_n))] + \frac{1}{D} \sum_{i=1}^{N-1} \sum_{j=i+1}^{N} [R_i R_j (\cos(\theta_i - \theta_j) + \cos(\varphi_i - \varphi_j) + \cos(\theta_i - \varphi_j) + \cos(\varphi_i - \theta_j))]. \quad (8)$$

symmetric elements by easily inverting the signal phase to obtain the symmetric matrix $\mathcal{S}_{N-1}$, can be given as follows:

$$\mathcal{S}_{N-1}=\boldsymbol{P}_{N-1}\boldsymbol{W}_{N-1}^H\mathcal{H}_{N-1}\boldsymbol{W}_{N-1}=\boldsymbol{U}_{N-1}\boldsymbol{V}_{N-1}\boldsymbol{Q}_{N-1}^H, \quad (15)$$

where the symmetric matrix $\mathcal{S}_{N-1}$ can be easily diagonalized, the Hermitian of $\boldsymbol{U}_{N-1}$ is decoding matrix, $\boldsymbol{Q}_{N-1}^H$ is the Hermitian of the precoding matrix $\boldsymbol{Q}_{N-1}$. We denote $\boldsymbol{V}_N$ as a diagonal block matrix with the element of the $k_N$th row as $\boldsymbol{V}_{N-1}$ and the remaining elements can be obtained from the $\mathcal{R}_N(k_N)$ ($k_N \in [0, K_N-1]$). In addition, the lowest-dimensional element of $\boldsymbol{V}_N$ corresponding to 1D transmit signal $s_{(l_N,\cdots,l_1)}$ can be denoted by $v_{(l_N,\cdots,l_1)}$, which denotes the singular values of equivalent channel matrix corresponding to $s_{(l_N,\cdots,l_1)}$.

Then, according to Eq. (15), we substitute Eq. (13) into Eq. (14) and futher write $\mathcal{R}_{N-1}$ as:

$$\begin{aligned}\mathcal{R}_{N-1} &= \boldsymbol{U}_{N-1}^H\boldsymbol{U}_{N-1}\boldsymbol{V}_{N-1}\boldsymbol{Q}_{N-1}^H\boldsymbol{Q}_{N-1}\boldsymbol{\Lambda}_{N-1}\tilde{\boldsymbol{S}}_{N-1} \\ &= \boldsymbol{V}_{N-1}\boldsymbol{\Lambda}_{N-1}\tilde{\boldsymbol{S}}_{N-1},\end{aligned} \quad (16)$$

we name the above-mentioned operation the $(N-1)$D OAM demodulation based on $(N-1)$D QF-UCA.

Then, performing a dimensionality reduction operation on Eq. (16). We take the $k_{N-1}$th row of $\mathcal{R}_{N-1}$ and denote as $\tilde{\boldsymbol{R}}_{N-2}$, which denotes the signal vector corresponding to the $(N-2)$D QF-UCA cells and can be given as follows:

$$\tilde{\boldsymbol{R}}_{N-2} = \boldsymbol{V}_{N-2}\boldsymbol{W}_{k_{N-2}}\tilde{\boldsymbol{S}}_{N-2} = \boldsymbol{V}_{N-2}\boldsymbol{W}_{N-2}\boldsymbol{\Lambda}_{N-2}\tilde{\boldsymbol{S}}_{N-2}. \quad (17)$$

The signal vector after the $(N-2)$D OAM demodulation based on $(N-2)$D QF-UCA, denoted by $\mathcal{R}_{N-2}$, can be derived as follows:

$$\mathcal{R}_{N-2}=\boldsymbol{W}_{N-2}^H\boldsymbol{V}_{N-2}^{-1}\boldsymbol{V}_{N-2}\boldsymbol{W}_{N-2}\boldsymbol{\Lambda}_{N-2}\tilde{\boldsymbol{S}}_{N-2}=\boldsymbol{\Lambda}_{N-2}\tilde{\boldsymbol{S}}_{N-2}. \quad (18)$$

Then, we progress to lower-dimensional QF-UCA demodulation. Performing a dimensionality reduction operation and taking the $k_2$th row of $\mathcal{R}_2$, we derive the signal vector $\tilde{\boldsymbol{R}}_1$ corresponding to 1D UCA cells can be given as follows:

$$\tilde{\boldsymbol{R}}_1 = \mathcal{R}_2(k_2) = \boldsymbol{W}_{k_1}\tilde{\boldsymbol{s}}_1 = \boldsymbol{W}_1\tilde{\boldsymbol{s}}_1. \quad (19)$$

The signal vector after the OAM demodulation based on 1D UCA, denoted by $\mathcal{R}_1$, can be derived as follows:

$$\mathcal{R}_1 = \boldsymbol{W}_1^H\tilde{\boldsymbol{R}}_1 = \boldsymbol{W}_1^H\boldsymbol{W}_1\tilde{\boldsymbol{s}}_1 = \tilde{\boldsymbol{s}}_1. \quad (20)$$

The SE of our $N$-dimensional OAM multiplex transmission scheme can be derived as follows:

$$SE = \sum_{l_N=0}^{K_N-1}\cdots\sum_{l_1=0}^{K_1-1}\log_2\left(1+\frac{\left|v_{(l_N,\cdots,l_1)}\right|^2\left|s_{(l_N,\cdots,l_1)}\right|^2}{\sigma^2(l_N,\cdots,l_1)}\right), \quad (21)$$

where $\sigma^2(l_N,\cdots,l_1)$ denotes the noise variance corresponding to the OAM-modes indexed with $(l_N,\cdots,l_1)$, $v_{(l_N,\cdots,l_1)}$ denotes the singular values of equivalent channel matrix corresponding to $s_{(l_N,\cdots,l_1)}$.

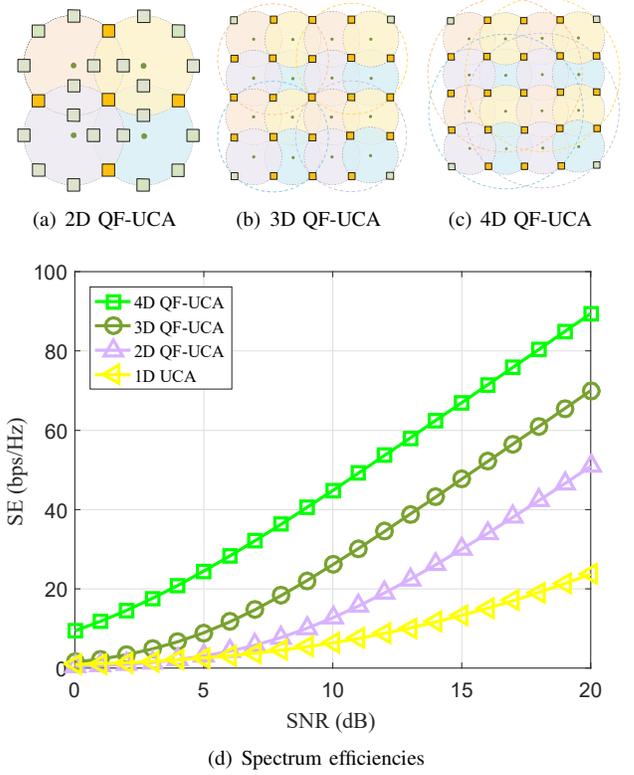

(a) 2D QF-UCA    (b) 3D QF-UCA    (c) 4D QF-UCA

(d) Spectrum efficiencies

Fig. 3. Spectrum efficiencies of the 4D, 3D, 2D, and 1D OAM multiplexing transmission as well as corresponding array-elements layouts (Fig. 3(a), Fig. 3(b), and Fig. 3(c) are the array-elements layouts of 2D, 3D, and 4D QF-UCA antennas consisting of 25 array-elements, respectively).

## IV. PERFORMANCE EVALUATIONS

In this section, numerical simulation results are presented to evaluate the performance of wireless communication for our developed $N$D OAM multiplexing transmission scheme. Firstly, we compared spectrum efficiencies of the 4D, 3D, 2D, and 1D OAM multiplexing transmission as well as corresponding array-elements layouts with a uniform array-elements number. Second, we compared spectrum efficiencies of the 4D, 3D, 2D, and 1D OAM multiplexing transmission for different transmission distances. Throughout the evaluation, we place the communication system in the 5.8 GHz frequency band, the constant $\beta$ is set to 1, and assume transceivers are aligned with each other.

Fig. 3 depicts the spectrum efficiencies of the 4D, 3D, 2D, and 1D OAM multiplexing transmission and corresponding array-elements layouts with a uniform array-elements number of 25. The antenna radius $R_E$ = 4m and the power is uniformly distributed across each OAM mode. As compared with the traditional single-loop UCA with 25 array-elements (referred to as 1D UCA), the higher-dimensional OAM multiplexing transmission corresponding to higher-dimensional QF-UCA achieve higher spectrum efficiencies. On the one hand, this is because OAM multiplexing transmission based on higher-dimensional QF-UCA antenna provide more orthogonal OAM-modes than the number of array-elements. On the other hand, the vortex electromagnetic wave of the higher-order OAM-modes diverge severely, and OAM multiplexing transmission based on higher-dimensional QF-UCA can utilize a greater number of lower-order OAM-modes, thus increasing SE.

In addition, a comparison between Fig. 3(b) and Fig. 3(c) reveals an intriguing observation: the array-elements layouts appear identical after design. However, Fig. 3(c) is a 4D QF-UCA with increased sharing of array-elements, facilitating the transmission of more orthogonal OAM-modes using the 4D OAM multiplexing transmission scheme. The higher-dimensional QF-UCA is different from the traditional single-loop UCA, and the diverse array-elements layout makes it possible for square arrays to generate vortex electromagnetic waves. Designing the layouts of array-elements and multiplexed transmission dimensions to facilitate the transmission of a greater number of orthogonal OAM-modes is an intriguing question that merits thorough investigation.

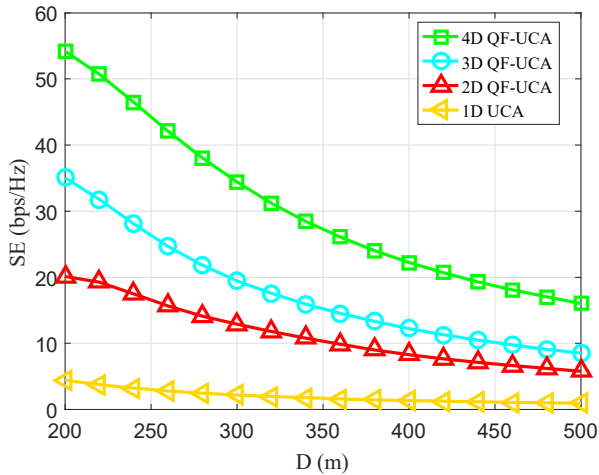

Fig. 4. Spectrum efficiencies of the 4D, 3D, 2D, and 1D OAM multiplexing transmission considering different transmission distances.

Fig. 4 depicts the spectrum efficiencies of the 4D, 3D, 2D, and 1D OAM multiplexing transmission considering different transmission distances. The array-elements layouts are consistent with the corresponding layouts in Fig. 3. The SNR = 15dB and the power is allocated on average to each OAM mode. As shown in Fig. 4, the traditional single-loop UCA with 25 array-elements (referred to as 1D UCA) exhibits lower SE compared to other higher-dimensional multiplexing transmission. This is because the higher-order OAM-modes of the 1D OAM multiplexing transmission diverge severely with distance, substantially decreasing the SE. However, higher-dimensional OAM multiplexing transmission based on the 4D QF-UCA can transmit more lower-order OAM-modes and still achieve relatively high spectrum efficiency at longer distances. We obtain that the higher-dimensional OAM multiplexing transmission based on higher-dimensional QF-UCA can significantly increase the transmission distance of the communication system without additional power and array-elements.

## V. Conclusions

In this paper, we proposed the $N$-dimensional OAM multiplex for high capacity transmission, which combines the design of antenna array-elements layout and multiplexing transmission scheme is a novel research direction to enhance the channel capacity. We design $N$-dimensional QF-UCA antenna structure and investigate the different array-elements layouts. Then, we developed the $N$-dimensional OAM modulation and demodulation schemes to transmit multiple OAM signals. Simulation results show that our proposed schemes can obtain a higher spectrum efficiency. The number of orthogonal data streams of the $N$-dimensional multiplexing transmission scheme far exceeds the number of antenna array-elements, providing a completely new solution to increase the number of available orthogonal streams.

## Acknowledgment

This work was supported in part by the National Key R&D Program of China under Grant 2021YFC3002102 and in part by the Key R&D Plan of Shaanxi Province under Grant 2022ZDLGY05-09.